\documentclass[12pt]{iopart}

\usepackage{graphicx} 
\usepackage{dcolumn} 
\usepackage{bm} 

\def\sqrtsNN{\mbox{$\sqrt{s_\mathrm{NN}}$}}

\newcommand{ \be }{\begin{equation}}    
\newcommand{ \ee }{\end{equation}}    
\newcommand{ \bea }{\begin{eqnarray}}    
\newcommand{ \eea }{\end{eqnarray}}    
\newcommand{ \la }{\langle}    
\newcommand{ \ra }{\rangle}

\usepackage{color}

\begin{document}       

\begin{flushright}    
\end{flushright}

\title{Directed and Elliptic Flow at RHIC}


\author{A.H. Tang\dag\ for the STAR Collaboration\footnote{For the full author list and acknowledgements see Appendix "Collaborations" in this volume.}}
\address{\dag\ NIKHEF and BNL \\ Physics Department, P.O. Box 5000, Brookhaven National Laboratory, Upton, NY~11973, aihong@bnl.gov}


\date{\today}

\begin{abstract}
We present the directed flow measurement ($v_1$) from Au+Au 
collisions at $\sqrtsNN = 62$ GeV. Over the pseudorapidity range
we have studied, which covers $\eta$ from $-1.2$ to $1.2$ and 
$2.4 < |\eta| < 4$,  the magnitude of $v_1$ for charged particles is 
found to increase monotonously with pseudorapidity for all centralities. 
No ``$v_1$ wiggle'', as predicted by various theoretical models, is observed 
at midrapidity. 
Elliptic flow ($v_2$) from moderate high $p_t$ particles ($3-6 GeV/c$) at 
$\sqrtsNN = 200$ GeV is presented as a function of impact parameter. 
It is found that models that are based on {\it jet quenching} alone 
appear to underpredict $v_2$ at moderate high $p_t$, while the model that 
incorporates both, recombination and fragmentation, describes the data better. 
\end{abstract}


\section{Introduction}\label{intro}

In non-central 
heavy ion collisions, the azimuthal distribution of emitted particles with 
respect to the reaction plane is not uniform. It can be 
characterized~\cite{methodPaper} by Fourier coefficients
\be
v_{n} = \la  \cos n (\phi - \psi) \ra
\label{vndef}
\ee
where $\phi$ denotes the azimuthal angle of an emitted particle, $\psi$ is
the orientation of the reaction plane, and $n$ denotes the harmonic.

The first Fourier coefficient, $v_1$, referred to as {\it directed flow},
describes the sideward motion of the fragments 
in ultra-relativistic nuclear collisions and it carries early 
information from the collision. Its shape at midrapidity is of special 
interest because it might reveal a signature of a possible phase transition 
from normal nuclear matter to a quark-gluon plasma~\cite{wiggle}.
Because of its importance, directed flow recently has attracted increased 
attention of both experimentalists and 
theoreticians~\cite{NA49,v1v4,MargueriteQM04,AihongQM04,
MarkusQM04,wiggle,HorstRBRC,LieWenChen}. In the paper~\cite{v1v4} that 
reports the first $v_1$ measurement at RHIC, the shape of $v_1$ 
at midrapidity is left ambiguous, due to the large statistical error.
It is now possible to answer this question 
with the large statistics obtained during RHIC run of 2004.

Elliptic flow ($v_2$) is caused by the initial geometric deformation of 
the reaction region in the transverse plane. At low transverse momentum, 
roughly speaking, large values of $v_2$ are considered signatures
of hydrodynamic behavior. At large transverse momentum, in a {\it jet quenching} 
picture~\cite{Wang01}, elliptic flow results from that jets emitted 
out-of-plane suffer more energy loss than those emitted in-plane. 
In an extreme case of {\it jet quenching}, particles are all emitted from the 
surface, as a consequence of that, $v_2$ is dominated by the geometry of the 
source. Thus it is interesting to study $v_2$ at large $p_t$, where 
the hydrodynamic description of the system is expected to break down and 
{\it jet quenching} is expected to happen, as a function of impact parameter.
Such study should give us a good constraint on various models.


\section{Directed flow}\label{v1}

The data for the $v_1$ analysis is based on the fourth year of operation of the
Relativistic Heavy Ion collider (RHIC) at $\sqrtsNN = 62.4$ GeV. 
The STAR detector~\cite{STAR} main Time Projection Chamber 
(TPC~\cite{STARTPC}) and two forward TPCs (FTPC~\cite{STARFTPC}) were 
used in the analysis. The data set consists of about 5 million   
minimum bias Au+Au events. The analysis is done with two methods, namely,
three-particle cumulant method~\cite{Borghini} and event plane method with 
mixed harmonics~\cite{MarkusQM04}.

\begin{figure}
  \begin{center}
\includegraphics[width=0.80\textwidth]{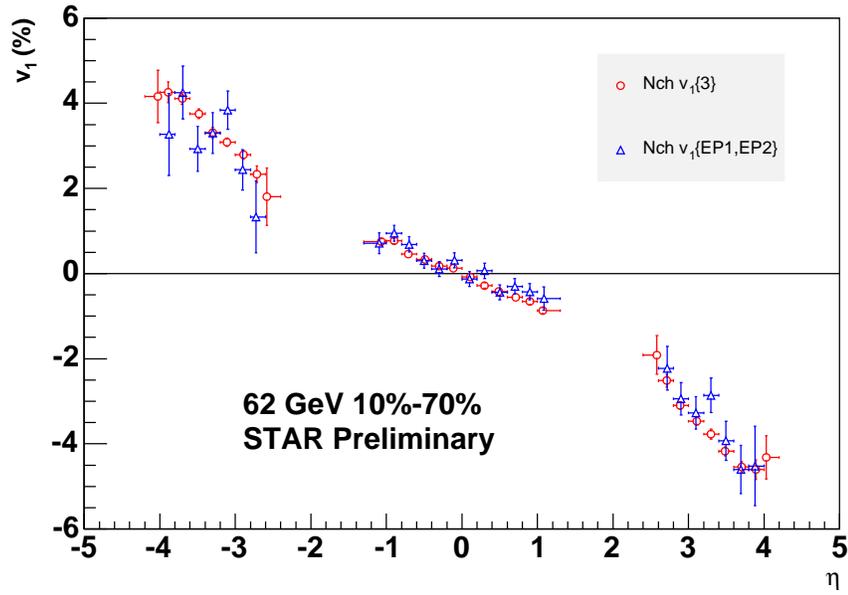}
\caption{ $v_1$ from three particle cumulant method (circles) and event plane method with mixed harmonics (triangles) as a function of pseudorapidity. Errors are statistical. The event plane result is from Markus Oldenburg.\label{fig:v1MarkusAndAihong}}
  \end{center}
\end{figure}

\begin{figure}
  \begin{center}
\includegraphics[width=0.80\textwidth]{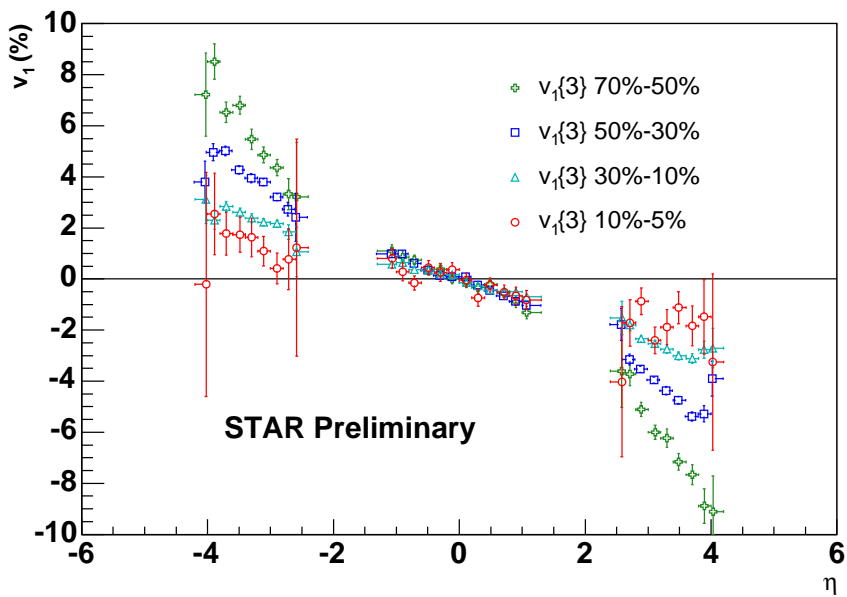}
\caption{ $v_1$ from three particle cumulant method as a function of pseudorapidity for four centrality bins. Errors are statistical. \label{fig:v1Eta4Cent}}
  \end{center}
\end{figure}

Fig.~\ref{fig:v1MarkusAndAihong} shows $v_1$ from the three-particle cumulants method
and the event plane method with mixed harmonics as a function of pseudorapidity ($\eta$). 
Both methods are based on three particle correlations and they should give the
same result, which is confirmed by the plot. The plot shows that over the 
pseudorapidity range we have studied, which covers $\eta$ from $-1.2$ to 
$1.2$ and $2.4 < |\eta| < 4$,  the magnitude of $v_1$ for charged particles 
is found increasing monotonically
with pseudorapidity for all centralities. No ``$v_1$ wiggle'' is observed at 
midrapidity as predicted by various theoretical models~\cite{wiggle}. The centrality 
dependence of $v_1$ is shown in Fig. ~\ref{fig:v1Eta4Cent}. As expected, in all
pseudorapidity regions, $v_1$ decreases with centrality. It is noticed that 
$v_1$ in the forward region decreases faster with centrality than that at 
midrapidity.

\begin{figure}
  \begin{center}
\includegraphics[width=0.80\textwidth]{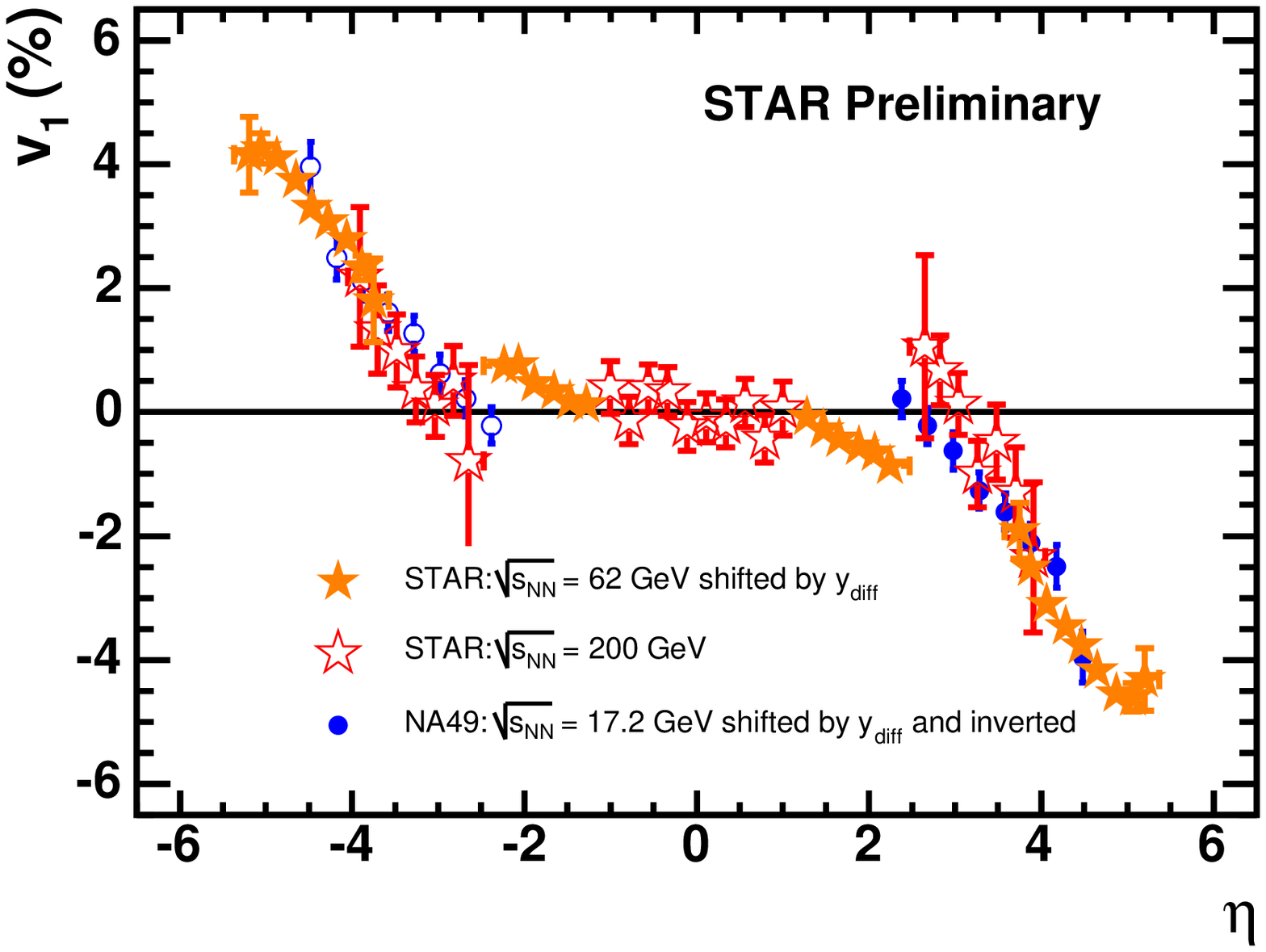}
\caption{ The values of $v_1$ from charged particles for 
Au+Au collisions at 200 GeV (open stars) and 62 GeV (solid stars) plotted as a function of pseudorapidity. Also shown are the results from NA49 (circles) for pions from 158{\it A} GeV Pb+Pb midcentral (12.5\%-33.5\%) collisions plotted as a function of rapidity. The open circles of NA49 have been reflected about midrapidity. The NA49 and 62 GeV points have been shifted plus or minus by the difference to 200 GeV in the beam rapidities. All results are from analyses involving 
three-particle cumulants, $v_1\{3\}$.\label{fig:v1Eta3Energies}}
  \end{center}
\end{figure}

Limiting fragmentation~\cite{LF} has successfully explained the spectra and some flow results 
in the forward region~\cite{PhobosSpectraAndV2,v1v4}. In Fig.~\ref{fig:v1Eta3Energies} we
show $v_1$ results from three different energies in the projectile 
frame relative to their respective beam rapidities. They look similar in the
forward region. 

\section{Elliptic flow}\label{v1}

The large value of the elliptic flow at high $p_t$~\cite{KirillHighptv2} 
and the strong suppression of back-to-back high $p_t$ jet-like 
correlations~\cite{STARBackToBack} support the jet-quenching
scenario qualitatively, however, 
the amount of elliptic flow observed at high $p_t$ for collisions at
\sqrtsNN = 130 GeV seems to exceed the values expected in the case of
complete quenching~\cite{Shuryak02}. 
Extreme quenching leads to emission of high-$p_t$ particles
predominantly from the surface, and in this 
case $v_2$ would be fully determined
by the geometry of the collision. 
This hypothesis can be tested by studying 
the centrality dependence of $v_2$ for high-$p_t$ particles.

\begin{figure}[t]       
  \includegraphics[width=0.8\textwidth]{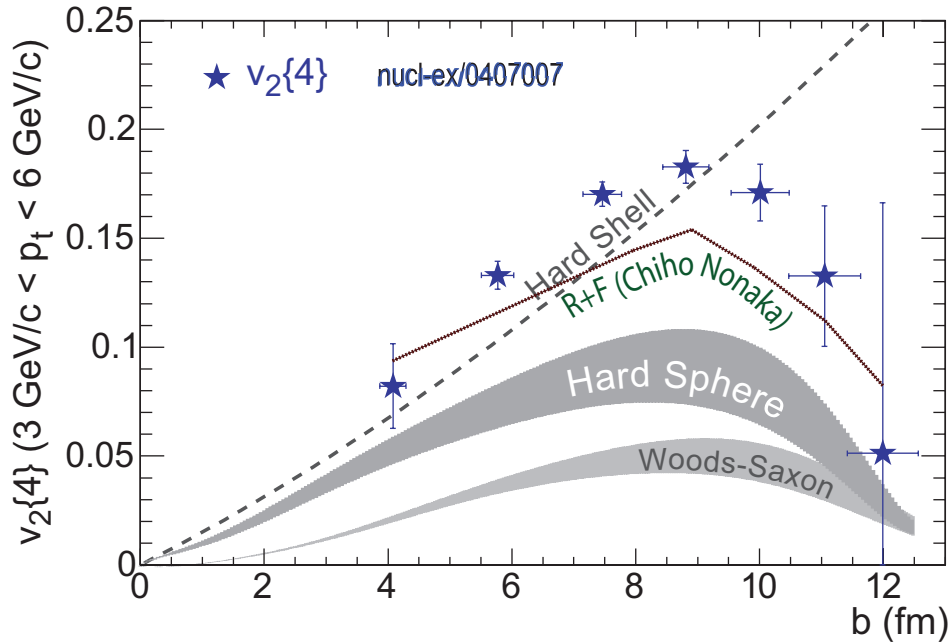}    
  \caption{
    $v_2$ at $3 \le p_t \le 6$ GeV/$c$ versus impact parameter, 
      $b$, compared to jet quenching models and the model that 
    incorporate both recombination and fragmentation. The data is
    from 200 GeV Au+Au collisions. See ref.~\cite{Chiho} for R+F
    calculation.
  }   
  \label{fig:v2SurfaceEmission_b}
\end{figure}

Figure~\ref{fig:v2SurfaceEmission_b} shows 
$v_2$ in the $p_t$-range of 3--6 GeV/$c$ (where $v_2$ is
approximately maximal and constant) versus impact parameter from Au+Au collision at $200$ GeV. For more description of the data set see ~\cite{highPtV204}.
The values of the impact parameters were obtained using 
a Monte Carlo Glauber calculation~\cite{Glauber}.  
The measured values of $v_2\{4\}$ are compared to various
simple models of jet quenching. 
The upper curve corresponds to a complete  
 quenching scenario, in which particles are emitted 
from a hard shell~\cite{SergeiQM02,Shuryak02};
this yields the maximum values of $v_2$ which are
possible under a surface emission assumption.
A more realistic calculation corresponds to a parameterization of jet
energy loss in a static medium where the absorption coefficient is set to
match the suppression of the inclusive hadron yields~\cite{Adams:2003kv}. 
The density distributions of the static medium are modeled using a
step function (following~\cite{XinNianJetQuenching03}) and a more 
realistic Woods-Saxon
distribution (following~\cite{Axel03}). The corresponding $v_2$ values are
shown as the upper and lower band, respectively. 
The lower and 
upper boundaries of bands correspond to an absorption 
that gives a suppression factor of 3 and 5~\cite{Adams:2003kv}, respectively, in central collisions.
Over the whole centrality range,  the measured $v_2$ values 
are much larger compared to calculations.  
Taking into account that this measurement is dominated by the yield at
lower $p_t$ ($3$ GeV/c),
the recombination of quarks might be responsible for the difference. Indeed
a model that combines the mechanism of both recombination and 
fragmentation~\cite{Chiho} gives a $v_2$ result that is larger than other 
models and is close to the data (see R+F curve in 
Figure~\ref{fig:v2SurfaceEmission_b}). It would be useful to have a quantitative 
estimate of the systematical uncertainty in this calculation so that the 
remaining discrepancy can be understood.

\section{Summary}\label{v1}


In summary, we have presented the $v_1$ measurement from Au+Au collisions of 
62 GeV at RHIC.  Over the pseudorapidity range
we have studied, which covers $\eta$ from $-1.2$ to $1.2$ and 
$2.4 < |\eta| < 4$, the magnitude of $v_1$ for charged particles is found 
to increase monotonically with pseudorapidity for all centralities. 
No ``$v_1$ wiggle'' for charged
particles, as predicted by various theoretical models, is observed at 
midrapidity. Viewed in the projectile
frame, $v_1$ from three different energies (17.2 GeV, 62.4 GeV and 200 GeV) 
looks similar, in support of limiting fragmentation hypothesis. We have studied
$v_2$ for moderate high $p_t$ particles from Au+Au collisions at 200 GeV, as
a function of centrality, and found that models that are based on 
{\it jet quenching} alone underpredict $v_2$. A model that combines both 
recombination and fragmentation describe the data better.


\Bibliography{99}


\bibitem{methodPaper}
  Poskanzer A.M. and Voloshin 1998
  {\it \PR c} {\bf 58} 1671

\bibitem{wiggle}
  Brachmann J \etal 2000
  {\it \PR C} {\bf 61} 024909  \\
  Bleicher M and St\"{o}cker H 2002
  {\it Phys. Lett. B} {\bf 526} 309  \\
  Csernai L P and Roehrich D 1999
  {\it Phys. Lett. B} {\bf 458} 454  \\
  Snellings R J M, Sorge H, Voloshin S A, Wang F Q and Xu N 2000
  {\it \PRL} {\bf 84} 2803


\bibitem{NA49}
  Alt C \etal [NA49~Collaboration] 2003
  {\it \PR C} {\bf 68} 034903

\bibitem{v1v4}
  Adams J. \etal [STAR Collaboration] 2004
  {\it \PRL} {\bf 92} 062301

\bibitem{MargueriteQM04}
  Tonjes M.B. \etal [PHOBOS Collaboration] 2004
  {\it J. Phys. G} {\bf 30} S1243-S1246

\bibitem{AihongQM04}
  Tang A.H. \etal [STAR Collaboration] 2004
  {\it J. Phys. G} {\bf 30} S1235-S1238 

\bibitem{MarkusQM04}
  Oldenburg M.D. \etal [STAR Collaboration]
  nucl-ex/0403007

\bibitem{HorstRBRC}
  Stoecker H.
  {\it Proceedings for RBRC Workshop ``New Discoveries  at RHIC''} nucl-th/0406018
  
\bibitem{LieWenChen}
  Chen L., Creco V., Ko C.M. and Kolb P.F.
  nucl-th/0408021
  
\bibitem{Wang01}
  Snellings R.J., Poskanzer A.M. and Voloshin S.A
  nucl-ex/9904003 \\
  Wang X.N. 2001
  {\it \PR C} {\bf 63} 054902

\bibitem{STAR}
  Ackermann K.H. \etal 2003
  {\it Nucl. Instrum. Method A} {\bf 499} 624

\bibitem{STARTPC}
  Anderson M. \etal 2003
  {\it Nucl. Instrum. Method A} {\bf 499} 659  

\bibitem{STARFTPC}
  Ackermann K.H. \etal 2003
  {\it Nucl. Instrum. Method A} {\bf 499} 713

\bibitem{Borghini}
  Borghini N, Dinh P M and Ollitrault J.-Y. 2002
  {\it \PR C} {\bf 66} 014905

\bibitem{LF}
  J.Benecke, T.T. Chou, C-N. Yang and E. Yen, {\it Phys. Rev.} {\bf 188}, 2159 (1969)

\bibitem{PhobosSpectraAndV2}
  Back B.B. \etal [PHOBOS Collaboration] 2003
  {\it \PRL} {\bf 91} 052303 \\
  Back B.B. \etal [PHOBOS Collaboration] 
  nucl-ex/0406012

\bibitem{KirillHighptv2}
  Adler C \etal STAR collaboration 2003
  {it \PRL} {\bf 90}, 032301

\bibitem{STARBackToBack} 
  Adler C \etal STAR collaboration 2003
  {\it \PRL} {\bf 90} 082302 

\bibitem{Shuryak02}        
  Shuryak E.V. 2002
  {\it \PR C} {\bf 66} 027902

\bibitem{highPtV204}
  Adams J. \etal [STAR Collaboration] 
  nucl-ex/0407007

\bibitem{Glauber}     
  Back B.B. \etal 2002
  {\it \PR C} {\bf 65} 031901(R) \\
  Adcox k. \etal 2001
  {\it \PRL} {\bf 86} 3500 \\
  Bearden I.G. \etal 2001
  {\it Phys. Lett. B} {\bf 523} 227 

\bibitem{SergeiQM02}        
  Voloshin S. A. 2003
  {\it Nucl. Phys. A} {\bf 715} 379c

\bibitem{Adams:2003kv}
  Adams J. \etal [STAR Collaboration] 2003
  {\it \PRL } {\bf 91} 172302

\bibitem{XinNianJetQuenching03} 
  Xin-Nian Wang, private communication. Calculation is based
 on the framework of nucl-th/0305010.

\bibitem{Axel03}        
  Drees A., Feng H. and Jia J.
  nucl-th/0310044

\bibitem{Chiho}        
  C. Nonaka, Talk given at {\it Hot Quark 2004} Conference, See this proceeding

\endbib
\end{document}